\documentclass[aps,preprint,floatfix,amsmath,superscriptaddress,endfloats,byrevtex]{revtex4}
\usepackage{graphicx}

\begin{document}

\title{Motion of quantum particle in dilute Bose-Einstein condensate at zero temperature}

\author{Alexey Novikov}
\affiliation{Department of Chemistry, University of Rochester,
P.O.Box 270 216,   Rochester NY 14627, USA}

\author{Mikhail Ovchinnikov}
\affiliation{Department of Chemistry, University of Rochester,
P.O.Box 270 216,   Rochester NY 14627, USA}

\date{\today}

\begin{abstract}
The motion of single quantum particle through Bose-Einstein
condensate (BEC) is considered within perturbation theory with
respect to the particle-BEC interaction. The Hamiltonian of BEC is
diagonalized by means of Bogoliubov's method. The process of
dissipation due to the creation of excitation in BEC is analyzed and
the dissipation rate is calculated in the lowest order of
perturbation theory. The Landau's criterion for energy dissipation
in BEC is then recovered. The energy spectrum of the impurity
particle due to the interaction with BEC and its effective mass are
evaluated.

\end{abstract}
%\pacs{03.65.Yz,05.30.-d,31.15.Kb}
% 03.65.Yz  Decoherence; open systems; quantum statistical methods
% 05.30.-d  Quantum statistical mechanics
% 05.40.-a Fluctuation phenomena, random processes, noise, and Brownian motion

%%%%%%%%%%%%%%%%%%%%%%
\maketitle
%
%%%%%%%%%%%%%%%%%%%%%%
\section{Introduction}
\label{intro}

Recent experiments on the rotational motion of molecules in the
superfluid helium droplets gave rise to a new interest in
superfluidity on a microscopic scale \cite{vilesov1998,
vilesov2000}. A unique properties of the measured rotational spectra
pose a large number questions, such as: what is the collective
molecule/superfluid wavefunction that describes sharp rotational
states observed in experiments; what are the properties of finite
systems and how is the limit of a bulk superfluid is reached
\cite{dumesh2006}?  A similar microscopic phenomena were studied in
the BEC of Sodium atoms in magnetic traps \cite{chikkatur2000}.  A
linear motion of impurities was shown to be dissipationless for the
speeds below the condensate speed of sound. A large number of
theoretical works described the molecule/droplet system using
imaginary time path integral Monte-Carlo approaches
\cite{dalfovo2001, whaley2004p, whaley2001k, whaley2004z,
whaley2004zk, whaley2003p, roy2006}. While in certain cases
remarkable agreement with experimental constants was obtained
\cite{roy2006}, those works are strictly limited to the calculation
of statistical properties and thus provides no real understanding of
the microscopic nature of the dissipationless motion.  The latter
can only be established by considering a real time dynamics. Several
theoretical works considered a motion of impurity through dilute
BEC. Recent works considered a macroscopic particle interacting with
dilute BEC \cite{suzuki2005}. In this case the motion of particle is
equivalent to the BEC in a time dependent external potential. This
problem was treated by solving time-dependent Gross-Pitaevskii
equations \cite{astrak2004}. A microscopic particle interacting with
the dilute BEC in Bogoliubov's approximation has been considered by
several authors using general Golden rule considerations
\cite{montina2003, montina2002, timmermans98}. These works were
based on the result of Miller et al \cite{miller62} which was
obtained using time-independent perturbation theory. The
Bogoliubov's treatment has also been successfully used for the
investigation of the force acting on the impurity particle due to
the quantum fluctuations in BEC \cite{roberts2006, roberts2005}.
Some authors treated a particle strongly interacting with
Bogoliubov's BEC and found a possibility of self localization
\cite{kalas2006, cucchietti2006, sacha2006}.

In this work we attempt to describe impurity moving through BEC as a
microscopic particle within time-dependent perturbation theory. We
introduce the Hamiltonian of a quantum particle moving within the
interacting Bose gas. No assumption is made about the relative mass
of an impurity compared to that of the Bose particles. We present
the analytic results for this problem in the limit of the dilute
Bose gas at zero temperature. After the introduction of the general
Hamiltonian, the Bogoliubov's approximation is made to convert the
Hamiltonian to the diagonal form.  Then the problem is reduced to
the quantum particle moving through the gas of non-interacting
Bogoliubov's excitations.  The initial conditions are chosen such
that an impurity is not correlated with the BEC.  The static
properties are achieved after the system reaches thermal equilibrium
in a long time limit of our treatment.

We obtain the expansion for the transition amplitude in powers of
particle-BEC interaction and restrict ourselves to the calculation
of the leading term of this expansion that corresponds to the Golden
rule approximation. The latter allows us to calculate various
observables such as: state-to-state transition probability,
dissipation rate, and the effective mass of the particle. Then we
obtain the criterion for the critical momentum of a particle below
which its motion becomes non-dissipative, i.e. the microscopic
version of the Landau's criterion. This criterion comes from the
exact momentum-energy conservation law. Finally, we calculate the
branch of the excitation spectrum corresponding to the moving
impurity, i.e. the energy of impurity as a function of its momentum.
Expanding the energy at small momenta it is found that the linear
part of energy vanishes and the impurity behaves like a free
particle with an effective mass that is determined by the second
order derivative of energy with respect to momentum.   We show that
our results agree with a time-independent treatment of Miller within
long time limit.

\section{MODEL HAMILTONIAN}
\label{model}

In this section the model for the interacting Bose gas is discussed
as well as the approximations that are made in our treatment. The
Hamiltonian of interacting Bose particles in coordinate
representation is given by
\begin{equation}
\label{ham1} H_B=-\sum_\alpha\frac{1}{2m}\frac{d^2}{d{\bf
r}_\alpha^2}+\frac{1}{2}\sum_{\alpha,\beta}U({\bf r}_\alpha-{\bf
r}_\beta)~.
\end{equation}
Here and throughout the paper we set $\hbar=1$. In the secondary
quantization the above Hamiltonian reads
\begin{equation}
\label{ham2} H_B=\sum_{\bf p}\frac{{\bf p}^2}{2m}b^+_{{\bf
p}}b_{{\bf p}}+\frac{1}{2V}\sum_{{\bf p}_1,{\bf p}_2,{\bf p}}U({\bf
p})b^+_{{\bf p}_1-{\bf p}}b^+_{{\bf p}_2+{\bf p}}b_{{\bf
p}_2}b_{{\bf p}_1}~,
\end{equation}
where $U({\bf p})$ is the Fourier transform of the interaction
potential
\begin{equation}
\label{potential} U({\bf p})=\int U({\bf r})e^{i{\bf pr}}d{\bf r}~.
\end{equation}
We will concentrate on the case of dilute gas $r_0\ll n^{-1/3}$
where $r_0$ is the range of potential on which $U(r)$ differs from
zero significantly and $n$ denotes density of gas. So the integral
(\ref{potential}) is significant in the domain $p_0\le 1/r_0$ while
characteristic momenta of particles are of the order of $p_c\le
n^{1/3}$. Thus due to diluteness condition $p_0\gg p_c$, and the
Fourier transform of the interaction $U({\bf p})$ in
Eq.~(\ref{ham2}) can be replaced by $U_0=\int U({\bf r})d{\bf r}$.
The zero Fourier component $U_0$ is connected with length of
$s$-scattering in first order Born approximation as following
\begin{equation}
a_s=\frac{m}{4\pi}U_0~.
\end{equation}
Let us consider degenerate Bose gas at zero temperature. In this
case the Hamiltonian ({\ref{ham2}) can be reduced to the diagonal
form with the help of the Bogoliubov's method \cite{bogol1947}. The
main assumption of the original treatment is that most of the
particles of the degenerate gas stay in the ground state with zero
energy and the number of particles with non-zero momentum $N_{{\bf
p}\not= 0}$ is much less then $N_0=N_{{\bf p}= 0}$. Thus the
creation/annihilation operators $b^+_{{\bf p}=0}$ and $b_{{\bf
p}=0}$ can be replaced by the macroscopic number $\sqrt{N_0}$. So
one can expand the interaction part of the Hamiltonian ({\ref{ham2})
in powers of $\sqrt{N_{{\bf p}\not= 0}/N_0}$ and leave the terms of
expansion up to second order only. After the described procedure the
Hamiltonian of the Bose gas attains the form
\begin{equation}
\label{ham3} H_B=\sum_{\bf p}\frac{p^2}{2m}b^+_{{\bf p}}b_{{\bf
p}}+\frac{U_0}{2V}\left [N^2-N+N\sum_{{\bf p}\not= 0}\left(b^+_{\bf
p}b^+_{\bf -p}+b_{\bf p}b_{\bf -p}+2b^+_{\bf p}b_{\bf
p}\right)\right ]~,
\end{equation}
where $N=N_0+N_{{\bf p}\not= 0}$. The next step is to diagonalize
the Hamiltonian (\ref{ham3}) introducing new bosonic
creation/annihilation operators $B^+_{\bf p}$ and $B_{\bf p}$
related to old operators $b^+_{\bf p}$ and $b_{\bf p}$ by linear
transformation
\begin{equation}
\label{transform} b_{\bf p}=\alpha_{\bf p}B_{\bf p}+\beta_{\bf
p}B^+_{\bf -p}~,~~~
\\
b^+_{\bf p}=\alpha_{\bf p}B^+_{\bf p}+\beta_{\bf p}B_{\bf -p}~.
\end{equation}
Substituting Eqs.~(\ref{transform}) into the Hamiltonian
(\ref{ham3}) and requiring the Hamiltonian to have diagonal form,
one gets for the transformation coefficients \cite{landau_v9}
\begin{eqnarray}
\alpha_{\bf p}&=&\frac{\mu_{\bf p}}{\sqrt{\mu_{\bf
p}^2-1}}~,~~~\beta_{\bf p}=\frac{1}{\sqrt{\mu_{\bf p}^2-1}}~,~~~
\\
\mu_{\bf p}&=&-\frac{\epsilon({\bf p})+p^2/2m+nU_0}{nU_0}~,\nonumber
\end{eqnarray}
together with the Hamiltonian in new representation
\begin{equation}
\label{ham4} H_B=E_0+\sum_{\bf p}\epsilon(p)B^+_{\bf p}B_{\bf p}~.
\end{equation}
The excitation spectrum $\epsilon(p)$ of the system yields
\begin{equation}
\epsilon(p)=\sqrt{\frac{p^2}{2m}\left(\frac{p^2}{2m}+2nU_0\right)}~,
\end{equation}
and has the phonon-like behavior at low momenta, i.e.
$\epsilon(p\to0)=p\sqrt{nU_0/m}=pc$, where $c$ is the speed of
sound. The ground state energy of BEC is given by
\begin{equation}
E_0=\frac{nU_0}{2}(N-1)+\frac{1}{2}\sum_{{\bf p}\not=
0}\left(\epsilon(p)-\frac{p^2}{2m}-nU_0\right)~.
\end{equation}

Now let us introduce a single quantum particle with mass $M$ and
coordinate ${\bf R}$ interacting with the environment of Bose gas
discussed above. The whole system is then described by following
Hamiltonian
\begin{equation}
\label{ham5} H=-\frac{1}{2M}\frac{d^2}{d{\bf R}^2}+H_B+H_{I}
\end{equation}
with the particle-environment interaction
\begin{equation}
\label{ham6} H_{I}=\frac{g}{V}\sum_{{\bf p},{\bf k}}b_{{\bf p}+{\bf
k}}^+b_{\bf p}e^{-i{\bf k}{\bf R}}~.
\end{equation}
Here the coupling constant $g$ is determined as zero Fourier
component of the system-environment interaction. In the spirit of
Bogoliubov's theory we expand the interaction Hamiltonian
(\ref{ham6}) in powers of $\sqrt{N_{{\bf p}\not= 0}/N_0}$ and leave
the terms of zeroth and first order, i.e. leading terms. Then the
Hamiltonian (\ref{ham6}) takes the form
\begin{equation}
\label{ham7}
H_{I}=gn+\frac{g\sqrt{N}}{V}\sum_{{\bf p}\not=
0}\left(b_{\bf p}^+e^{-i{\bf p}{\bf R}}+b_{{\bf p}^\prime}e^{i{\bf
p}{\bf R}}\right)~.
\end{equation}
We neglected the terms of the order of $N_{p\not=0}$ in
(\ref{ham7}). As it becomes clear below, the contribution of the
higher order terms to the dissipation process is of the order of
$N_{{\bf p}\not=0}$ which would be an excess precision compared with
the Bogoliubov's approximation made above. Expressing operators
$b_{\bf p}^+$ and $b_{\bf p}$ through new operators $B_{\bf p}^+$
and $B_{\bf p}$, and using the secondary quantization representation
for the system particle, for the full Hamiltonian (\ref{ham5}) one
gets
\begin{eqnarray}
\label{fullham}
H&=&E^\prime_0 +\sum_{\bf q}\frac{{\bf
q}^2}{2M}a_{\bf q}^+a_{\bf q}+\sum_{\bf p}\epsilon(p)B^+_{\bf
p}B_{\bf p}+\sum_{{\bf q},{\bf p}\not= 0}\gamma_{\bf
p}\left(a^+_{{\bf q}-{\bf p}}a_{\bf q}B^+_{\bf p}+a^+_{{\bf q}+{\bf
p}}a_{\bf q}B_{\bf p}\right)~,
\\
\gamma_{\bf
p}&=&\frac{g}{V}\sqrt{\frac{Np^2}{2m\epsilon(p)}}~.\nonumber
\end{eqnarray}
Here operators $a^+_{\bf q}/a_{\bf q}$ create/annihilate the
particle in state $|{\bf q}\rangle$, and the ground state energy
$E_0^\prime=E_0+gn$ is shifted with respect to the $E_0$ due to the
system-condensate interaction.

\section{TRANSITION PROBABILITY AND DISSIPATION RATE}
\label{transit} The next step is the calculation of the transition
probability for the whole system from some initial state $|i\rangle$
to final state $|f\rangle$. So our aim is to construct the
perturbative expansion for the matrix element of the evolution
operator  $a_{if}(t)=\langle f|e^{-iHt}|i\rangle$ in powers of
particle-BEC interaction (\ref{ham7}). Here we restrict ourselves to
the calculation of the leading, i.e. lowest order term of this
expansion only. Let us write  the matrix element of evolution
operator in interaction representation
\begin{eqnarray}
a_{if}(t)&=&\langle f|e^{-iH_0t}{\rm
T}\exp\left(-i\int_0^tH_I(s){\rm d}s\right)|i\rangle~,
\\
H_0&=&H-H_I~,~~~H_I(s)=e^{iH_0s}H_Ie^{-iH_0s}~.
\end{eqnarray}
Initial and final states of the system particle are supposed to be
free particle states $|{\bf q}\rangle=\frac{1}{\sqrt{V}}e^{i{\bf
q}{\bf R}}$ with momenta ${\bf q}_i$ and ${\bf q}_f$, respectively.
The Bose environment is initially in its ground state $|0\rangle_B$
and then evolutes into some final state $|f\rangle_B$. Note that the
state $|0\rangle_B$ is the ground state of the diagonalized
Hamiltonian (\ref{ham4}) with the ground state energy $E^\prime_0$
and it does not equal the state in which all of the Bose particles
have zero momentum. Below we will find the transition amplitude as
an expansion in powers of coupling constant $g$, i.e.
\begin{equation}
a_{if}(t)=a_{if}^{(0)}(t)+a_{if}^{(1)}(t)+a_{if}^{(2)}(t)+...~,~~~
a_{if}^{(n)}(t)\sim g^n~.
\end{equation}
The zeroth order term is given by $a_{if}^{(0)}(t)=\delta_{{\bf
q}_f,{\bf q}_i}\exp\left[-i(E_0^\prime-q_i^2/2M)t\right]$. The first
order term of expansion describes the process of creation of
excitation in BEC with the momentum ${\bf p}_f$ due to the
interaction with the particle
\begin{eqnarray}
a_{if}^{(1)}(t)=-i\int_0^tds\sum_{{\bf p},{\bf q}}\gamma_{\bf
p}\langle{\bf p}_f|_B\langle{\bf q}_f|e^{-iH_0(t-s)}a^+_{{\bf
q}-{\bf p}}a_{\bf q}B^+_{\bf p}e^{-iH_0s}|{\bf
q}_i\rangle|0\rangle_B \nonumber
\\
=-i\gamma_{{\bf p}_f}\delta_{{\bf q}_f,{\bf q}_i-{\bf
p}_f}\int_0^tds\exp\left[-i\left(E_0^\prime-
q_i^2/2M\right)t-i\omega({\bf p})(t-s)\right]~,
\end{eqnarray}
where $\omega({\bf p})=\epsilon(p)+p^2/2M-{\bf q}_i{\bf p}/M~$. The
term of second order is defined by two processes: first, is creation
of two excitation in BEC; and second, is creation and annihilation
of one excitation in BEC while the particle finally remains in its
initial state. We will consider the amplitude of the latter process
since only this amplitude gives the contribution of second order
into the transition probability. Thus the second order matrix
element reads
\begin{eqnarray}
a_{if}^{(2)}(t)=-\frac{1}{2}\int_0^tds\int_0^tds\prime\langle
0|_B\langle{\bf q}_f|e^{-iH_0t}{\rm
T}\left(H_I(s)H_I(s^\prime)\right)|{\bf q}_i\rangle|0\rangle_B
\nonumber
\\
=-\delta_{{\bf q}_f,{\bf q}_i}\sum_{\bf
p}\int_0^tds\int_0^tds^\prime\Theta(s-s^\prime)\exp\left[-i\left(E_0^\prime-
q_i^2/2M\right)t-i\omega({\bf p})(s-s^\prime)\right]~.
\end{eqnarray}
Finally, one can find the probability of the particle transition
from state $|{\bf q}_i\rangle$ to $|{\bf q}_f\rangle$ in the
following form
\begin{equation}
\label{pr} w_{if}=\delta_{{\bf q}_i,{\bf q}_f}\left(1-\sum_{\bf
p}\gamma_{\bf p}^2\left|\int_0^tdse^{-is\omega({\bf
p})}\right|^2\right)+\gamma^2_{{\bf q}_i-{\bf
q}_f}\left|\int_0^tdse^{-is\omega({\bf q}_i-{\bf q}_f)}\right|^2~.
\end{equation}
The standard way to deal with the integral $\int_0^tdse^{-i\omega
s}$ in the expression for the transition probability (\ref{pr}) is
to evaluate its value in the limit of long time interval
$t\gg1/\omega$. So for the square of the absolute value of this
integral one gets
\begin{equation}
\label{limit1}
\left|\int_0^tdse^{-is\omega({\bf
p})}\right|_{t\to\infty}^2\longrightarrow ~2\pi t\delta(\omega({\bf
p}))~.
\end{equation}

First, let us discuss the probability of the particle of leaving its
initial state $|{\bf q}_i\rangle$  and moving to the state with a
different arbitrary momentum $|{\bf q}_f\rangle$ in a unit of time
with creation of the Bogoliubov's excitation in the surrounding BEC
with the momentum ${\bf p}={\bf q}_f-{\bf q}_i$. In the long time
approximation the probability reads
\begin{equation}
dw({\bf p})=2\pi\gamma_{\bf p}^2\delta(\omega({\bf p}))dt~.
\end{equation}
So the maximum momentum $p_M$ that can be transmitted from particle
to BEC is collinear to ${\bf q}_i$ and determined by the crossing
point of the line $q_ip/M$ and the curve $\epsilon(p)+p^2/2M$
\begin{equation}
\label{p0}
p_M=\frac{2}{1-(M/m)^2}\left(q_i-\sqrt{q_c^2+(M/m)^2(q_i^2-q_c^2)}\right),~~~q_c=Mc~.
\end{equation}
Besides, the maximum angle $\theta_M$ between vectors ${\bf q}_i$
and ${\bf p}$ is given by the condition $\cos\theta_M=q_c/q_i$. Thus
we have the Cherenkov like emission of the BEC excitation with the
length of the momentum $0<|{\bf p}|<p_M$ into the cone with the
angle $\theta_m$ along the vector ${\bf q}_i$.

Next, we can define the transition rate as the probability of the
particle to leave its initial state $|{\bf q}_i\rangle$ in a unit of
time as
\begin{equation}
\label{gamma0}
\Gamma_T=\sum_{\bf p}\frac{dw(\omega({\bf
p}))}{dt}
%=2\pi\sum_{\bf p}\gamma_{\bf p}^2\delta(\omega({\bf p}))~.
\end{equation}
In the thermodynamic limit ($N\to\infty,~~V\to\infty,~~N/V=n$) one
has to replace the sum over momenta $\sum_{\bf p}$ by the integral
$\frac{V}{(2\pi)^3}\int d{\bf p}$ in the right hand side of
Eq.~(\ref{gamma0}), and for the transition rate one gets
\begin{equation}
\label{gamma1} \Gamma_T=\frac{ng^2}{8\pi^2m}\int d{\bf
p}\frac{p^2}{\epsilon(p)}\delta(\omega({\bf p}))~.
\end{equation}
Let us write the integral in Eq.~(\ref{gamma1}) in the following
form
\begin{equation}
\label{gamma2} \Gamma_T=\frac{nMg^2}{4\pi mq_i}\int_0^\infty
dp\frac{p^3}{\epsilon(p)}\int_0^1dx\delta(x-x_0)~,
\end{equation}
where $x={\bf p}{\bf q}_i/pq_i$ and the value $x_0$ is determined
from the condition $\omega({\bf p})=0$. So the integral over $x$ is
unit only if $x_0<1$, or
\begin{equation}
\label{cond} \epsilon(p)+p^2/2M<pq_i/M~.
\end{equation}
Clearly the condition (\ref{cond}) is never fulfilled if the
absolute value of the initial momentum of the particle $q_i$ is less
then $M\epsilon(p\to0)/p=Mc=p_c$, and then the integral in the
expression for the transition rate $\Gamma_T$ is always zero. At
this point we recovered the Landau's criterion - there is no energy
dissipation if the speed of particle $v_i=q_i/M$ moving through the
interacting Bose gas is less than speed of sound.

To evaluate the rate for momenta above the critical one we proceed
further with the integral (\ref{gamma2})
\begin{equation}
\label{gamma3} \Gamma_T=\frac{nMg^2}{4\pi
mq_i}\int_0^{p_M}dp\frac{p^3}{\epsilon(p)}~.
\end{equation}
Here the integration is performed over momentum $p$ from zero to the
crossing point $p=p_M$ given by Eq.~(\ref{p0}). Finally, the
transition rate reads
\begin{equation}
\label{gammaf} \Gamma_T=\frac{Mmng^2}{2\pi
q_i}\left[\epsilon(p_M)-mc^2\ln\left(1+\frac{\epsilon(p_M)+p_M^2/2m}{mc^2}\right)\right]~.
\end{equation}
Let us consider the limit case when the initial momentum $q_i$ is
close to its critical value $q_c$. Expanding the expression
(\ref{p0}) in powers of small $q_i-q_c$ and assuming the value $M/m$
to be finite, for the upper limit of integration we obtain
$p_M\approx2(q_i-q_c)$. Then expanding the right hand side of
Eq.(\ref{gammaf}) in powers of $p_M/mc$, for the leading term of
expansion one gets
\begin{equation}
\Gamma_T(q_i\sim q_c)=\frac{2ng^2}{3\pi mc^2}(q_i-q_c)^3~.
\end{equation}
So the lowest term of expansion appears to be of the third order of
small value $q_i-q_c$. In the opposite case of large initial
momentum $q_i\gg q_c$ one can obtain the following expression for
the transition rate
\begin{equation}
\Gamma_T=\frac{ng^2Mq_i}{\pi}\left(\frac{m}{M+m}\right)^2~,
\end{equation}
which is surely independent on speed of sound.

Our next example is the calculation of the energy transferred from
particle to BEC as a function of time
\begin{equation}
E(t)=\sum_{\bf p}\epsilon(p)\gamma^2_{\bf
p}\left|\int_0^tdse^{-is\omega({\bf p})}\right|^2_{t\to\infty}=2\pi
t\frac{V}{(2\pi)^3}\int d{\bf p}\gamma_{\bf
p}^2\epsilon(p)\delta(\omega({\bf p}))~.
\end{equation}
Here we can determine the dissipation rate as the amount of energy
transferred in unit of time
\begin{equation}
\label{gammae} \Gamma_E=\frac{dE(t)}{dt}=\frac{ng^2}{8\pi^2m}\int
d{\bf p}p^2\delta(\omega({\bf p}))~.
\end{equation}
Calculating the integral in (\ref{gammae}) in the same manner as
described above, for the dissipation rate one gets
\begin{equation}
\label{gammae1} \Gamma_E=\frac{Mng^2p_M^4}{16\pi mq_i}~.
\end{equation}
In the limit case of massive particle $M\gg m$ the expression
(\ref{gammae1}) converges to the following value
\begin{equation}
\Gamma_E=\frac{ng^2(q_i^2-q_c^2)^2}{\pi
q_i}\left(\frac{m}{M}\right)^3~.
\end{equation}
This result exactly corresponds to the rate of energy dissipation
obtained in \cite{suzuki2005} for the classical particle moving
through BEC with constant velocity.

\section{PARTICLE ENERGY SPECTRUM AND EFFECTIVE MASS}

In this section we will consider the non-dissipative motion of the
impurity though the BEC when the initial momentum of the impurity
particle $q_i$ is less than the critical momentum $q_c=Mc$ at which
the dissipation process occurs according to the Landau's criterium.
In the absence of dissipation one would expect that the impurity
will move as a free particle but its energy should be shifted due to
the coupling with the BEC environment. Thus our purpose is to
calculate the contribution to the energy of the impurity due to the
interaction with BEC and the effective mass of the particle.

Let us start with the calculation of the full energy of the
particle-BEC system as a function of time. Following the perurbative
treatment developed above, one can write the energy of the system as
an expansion in powers of coupling parameter $g$
\begin{equation}
E(t)=E^{(0)}(t)+E^{(1)}(t)+E^{(2)}(t)~,
\end{equation}
coming from the expansion for the evolution of density matrix
\begin{equation}
\rho(t)=\rho^{(0)}(t)+\rho^{(1)}(t)+\rho^{(2)}(t)~.
\end{equation}
The system energy as a function of time is then given by $E(t)={\rm
Tr}\left(H\rho(t)\right)$, where $H$ is the Hamiltonian of the full
system (\ref{fullham}) and ${\rm Tr}$ denotes the trace over all
possible states. The zeroth order contribution corresponding to the
unperturbed system is
\begin{equation}
E^{(0)}={\rm
Tr}\left(H_0\rho^{(0)}\right)=E_0^\prime+\frac{q_i^2}{2M}~.
\end{equation}
The first order does not bring any contribution since ${\rm
Tr}\left(H_0\rho^{(1)}(t)+H_I\rho^{(0)}(t)\right)=0$. The second
order term consists of two terms $E^{(2)}={\rm
Tr}\left(H_0\rho^{(2)}\right)+{\rm Tr}\left(H_I\rho^{(1)}\right)$.
Let us consider the first term
\begin{equation}
\label{term1}
{\rm
Tr}\left(H_0\rho^{(2)}\right)=\sum_{f}w^{(2)}_{i,f\not=i}\left(\frac{q_f^2}{2M}+\epsilon(p_f)+E_0^\prime\right)+
w^{(2)}_{i,f=i}\left(\frac{q_i^2}{2M}+E_0^\prime\right)~,
\end{equation}
where the transition probability $w_{if}^{(2)}$ is defined by
Eq.~(\ref{pr}) as $w^{(2)}_{if}=w_{if}-1$. We are interested again
in the long time limit when the whole system reaches its equilibrium
state, so the upper limit of integration over time in the right hand
side of Eq.~(\ref{pr}) must be replaced by infinity. Since here we
consider the case of dissipationless motion of the impurity,
$\omega({\bf p})\not= 0$ and instead of the limit (\ref{limit1}) one
has to use following formula
\begin{equation}
\left|\int_0^tdse^{-is\omega({\bf
p})}\right|_{t\to\infty}^2\longrightarrow \frac{1}{\omega^2({\bf
p})}~.
\end{equation}
Hence for the term (\ref{term1}) we have
\begin{equation}
{\rm Tr}\left(H_0\rho^{(2)}\right)=\sum_{\bf p}\frac{\gamma^2_{\bf
p}}{\omega({\bf p})}~.
\end{equation}
The remaining term ${\rm Tr}\left(H_I\rho^{(1)}\right)$ is given by
\begin{equation}
{\rm Tr}\left(H_I\rho^{(1)}\right)=\sum_{f\not=i}\left(\langle
f|H_I|i\rangle\langle i|\rho^{(1)}(t)|f\rangle~+~c.c. \right)~.
\end{equation}
Using the expression for the interaction Hamiltonian
Eq.~(\ref{ham7}) and noting that $\langle
i|\rho^{(1)}(t)|f\rangle=a_{if}^{(0)}(t)a_{if}^{(1)*}(t)$, for the
above term one gets
\begin{equation}
\label{term2}
{\rm Tr}\left(H_I\rho^{(1)}\right)=-2\sum_{\bf
p}\gamma_{\bf p}^2\int_0^tds\sin(s\omega({\bf p}))~.
\end{equation}
In the long time limit $\int_0^\infty ds\sin(s\omega({\bf
p}))={\mathcal P}\frac{1}{\omega({\bf p})}$, where ${\mathcal P}$
denotes principal value. Finally, the energy shift given by the sum
of terms (\ref{term1}) and (\ref{term2}) reads
\begin{equation}
\label{en11} E^{(2)}=-\sum_{\bf p}\frac{\gamma^2_{\bf
p}}{\omega({\bf p})}=-\frac{ng^2}{16\pi^3m}\int d{\bf
p}\frac{p^2}{\epsilon(p)\omega({\bf p})}~.
\end{equation}
As was mentioned above, here we are interested in evaluation of the
energy of the impurity particle in case of relatively small initial
momentum $q_i$. In this case the energy can be written as an
expansion in powers of small parameter $q_i/q_c$ up to second order,
i.e.
\begin{equation}
\label{en12}
E_p(q_i)=E_p(0)+\frac{q_i^2}{2M_{ef}}~.
\end{equation}
It turns out that the term linear in $q_i$ disappears, and the
effective mass of the particle in (\ref{en12}) is determined as
follows
\begin{equation}
\frac{1}{M_{ef}}=\frac{1}{M}+\left.\frac{d^2E^{(2)}}{dq_i^2}\right|_{q_i=0}~.
\end{equation}

Let us start with the calculation of the zeroth order contribution
in powers of $q_i$
\begin{equation}
\label{en1}
E_p(0)=ng-\frac{1}{4\pi^2}\frac{ng^2}{m}\int_0^\infty\frac{p^4dp}{\epsilon(p)\big(\epsilon(p)+p^2/2M\big)}~.
\end{equation}
The integral in the right hand side of Eq.~(\ref{en1}) diverges at
large momentum like $4mm_r\int_0^\infty dp$, where
$m_r=\left(1/m+1/M\right)^{-1}$ is reduced mass of the system of the
relevant and the BEC particles. The standard way to avoid the such
kind of divergence is to renormalize the coupling constant
introducing some another expansion parameter. Let us determine the
coupling parameter $g$ as an expansion in powers of scattering
length $a$
\begin{equation}
\label{reexpansion} g=\frac{2\pi a}{m_r}(1+\kappa a+...)~,
\end{equation}
and then remove the divergence from the expansion for the energy
(\ref{en1}) to the expansion (\ref{reexpansion}). Substituting
equation (\ref{reexpansion}) into Eq.~(\ref{en1}), leaving terms up
to second order in powers of scattering length $a$ and requiring the
energy $E_p(0)$ to be finite, for the renormalized coupling constant
one gets
\begin{equation}
\label{reexpansion1} g=\frac{2\pi
a}{m_r}\left(1+\frac{2a}{\pi}\int_0^\infty dp\right)~.
\end{equation}
The actual physical reason of the divergence is that the zero
Fourier component of the interaction is used as the coupling
constant $g$. The procedure of renormalization of coupling constant
corresponds to taking into account the second order Born
approximation and can also be done employing direct expansion of the
scattering length in powers of coupling constant \cite{landau_v9}.
Using the relation (\ref{reexpansion1}), the energy $E_p(0)$ is now
finite and reads
\begin{eqnarray}
\label{en2} E_p(0)&=&\frac{2\pi
na}{m_r}\left[1+\frac{4amc}{\pi}I_0\left(\frac{m}{M}\right)\right]~,
\\
I_0(x)&=&\frac{x\sqrt{x^2-1}-\ln(x+\sqrt{x^2-1})}{(x-1)\sqrt{x^2-1}}~.\nonumber
\end{eqnarray}

\begin{figure}
\includegraphics[width=7.0cm]{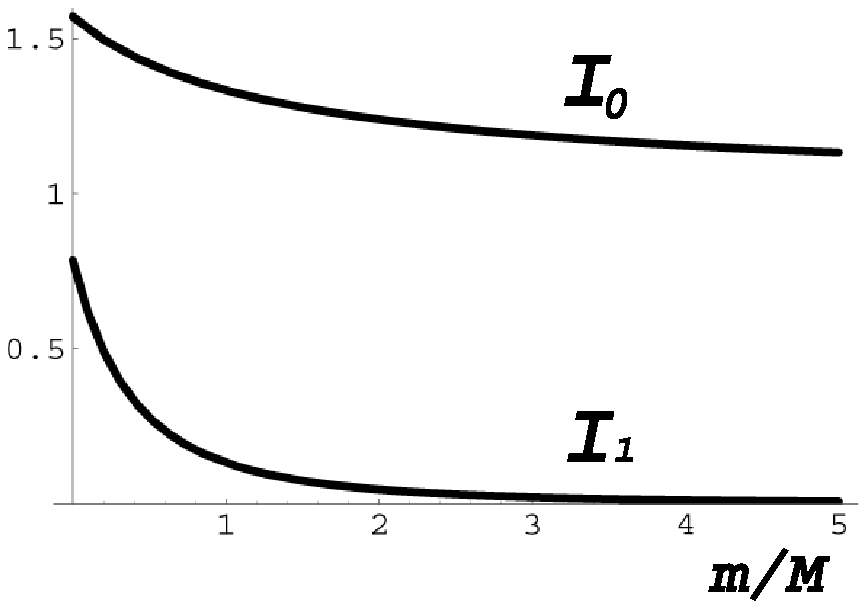}
\caption{ \label{fig} Integrals $I_0$ and $I_1$ as the functions of
relation $m/M$~. }
\end{figure}
Evaluating the term of the second order expansion in the expression
for the energy (\ref{en11})
\begin{equation}
\label{en3}
\left.\frac{d^2E^{(2)}}{dq_i^2}\right|_{q_i=0}=-\frac{2}{3}\frac{na^2}{mm_r^2M^2}
\int_0^\infty\frac{p^6dp}{\epsilon(p)\big(\epsilon(p)+p^2/2M\big)^3}~,
\end{equation}
for the effective mass of the particle we have
\begin{equation}
\label{mef}
M_{ef}=M\left[1-\frac{32}{3}\frac{na^2}{Mc(m_r/m)^2}I_1\left(\frac{m}{M}\right)\right]^{-1}~,
\end{equation}
where
\begin{equation}
I_1(y)=\frac{1}{2(y^2-1)^{5/2}}\left[(1+2y^2)\ln(y+\sqrt{y^2-1})-3y\sqrt{y^2-1}\right]~.
\end{equation}
The behavior of the functions $I_0$ and $I_1$ in expressions for the
energy (\ref{en2}) and the effective mass (\ref{mef}), respectively,
is shown in Fig.~(\ref{fig}). In case of heavy particle $M\gg m$ its
effective mass yields $M_{ef}=M(1-8\pi na^2/3Mc)^{-1}$, the same
result as was obtained in \cite{navez1999}.

\section{CONCLUSION}
\label{conclusion}

The goal of this paper is the understanding of the effect of
superfluidity reflected on a motion of a microscopic impurity inside
a BEC environment.  Quantitatively our system is very different from
the realistic liquid helium. However, for dilute BEC we obtained
physical behavior that gives qualitative understanding of
manifestation superfluidity on a microscopic scale.  We calculated
the transition amplitude to the second order of the perturbation
theory.  The inelastic contribution to this amplitude responsible
for the dissipation comes from the first order of the perturbation
expansion. It is shown that at the velocities less than speed of
sound this term is zero and we recover the exact Landau's criterion.
The elastic part corresponds to the second order term and is
responsible for the shift of the particle energy. Divergence of this
energy shift which appears because of the delta-functional potential
is avoided by renormalization of coupling constant.  The energy of
the particle is calculated as a function of its momentum. At small
momenta this function is quadratic, thus the impurity moves like a
free particle but its mass is shifted due to the interaction with
the BEC. The natural extension of this work is the calculation of
the Green's function of an impurity and the development of the
perturbation theory for its poles.  This work is currently in
progress.

\section{ACKNOWLEDGEMENTS}

The support for this work has been provided by the NSF (CAREER award
no. 0645340) and by the University of Rochester.


\begin{thebibliography}{10}

\bibitem{landau_v9}
E.~M. Lifshitz and L.~P. Pitaevskii,
\newblock Statistical Physics, Part 2,
\newblock Pergamon Press, Oxford, 1980.


\bibitem{bogol1947}
N.~N. Bogoliubov,
\newblock J. Phys. (Moscow) 11 (1947) 23

\bibitem{suzuki2005}
Jun Suzuki,
\newblock arXiv: cond-mat/0407714 (2005).

\bibitem{astrak2004}
G.~E. Astrakharchik and L.~P. Pitaevskii,
\newblock Phys. Rev. A 70 (2004) 013608.

\bibitem{navez1999}
P.~Navez and M.~Wilkens,
\newblock J. Phys. B: At. Mol. Opt. Phys. 32 (1999) L629.

\bibitem{vilesov1998}
S.~Grebenev, P.~Toennis and A.~Vilesov,
\newblock Science 279 (1998) 2083.

\bibitem{vilesov2000}
S.~Grebenev, B.~Sartakov, P.~Toennis and A.~Vilesov,
\newblock Science 289 (2000) 1532.

\bibitem{dumesh2006}
B.~S.~Dumesh and L.~A.~Surin,
\newblock Physics-Uspekhi 49 (2006) 11

\bibitem{dalfovo2001}
F.~Dalfovo and S.~Stringary,
\newblock J. Chem. Phys. 115 (2001) 10078

\bibitem{whaley2004p}
F.~Paesani and K.~B.~Whaley,
\newblock J. Chem. Phys. 121 (2004) 5293

\bibitem{whaley2001k}
Y.~Kwon and K.~B.~Whaley,
\newblock J. Chem. Phys. 114 (2001) 3163

\bibitem{whaley2004z}
R.~E.~Zillich and K.~B.~Whaley,
\newblock Phys. Rev. B 69 (2004) 104517

\bibitem{whaley2004zk}
R.~E.~Zillich, Y.~Kwon and K.~B.~Whaley,
\newblock Phys. Rev. Lett. 93 (2004) 250401

\bibitem{whaley2003p}
M.~V.~Patel, A.~Viel, F.~Paesani, P.~Huang and K.~B.~Whaley,
\newblock J. Chem. Phys. 118 (2003) 5011

\bibitem{roy2006}
W.~Topic, W.~Jaeger, N.~Blinov, P.-N.~Roy, M.~Botti, and S.~Moroni,
\newblock J. Chem. Phys. 125 (2006) 144310


\bibitem{kalas2006}
R.~Kalas and D.~Blume,
\newblock Phys. Rev. A 73 (2006) 043608


\bibitem{cucchietti2006}
F.~M.~Cucchietti and E.~Timmermans,
\newblock Phys. Rev. Lett. 96 (2006) 210401

\bibitem{sacha2006}
K.~Sacha and E. Timmermans,
\newblock Phys. Rev. A 73 (2006) 063604

\bibitem{montina2003}
A.~Montina,
\newblock Phys. Rev. A 67 (2003) 053614


\bibitem{montina2002}
A.~Montina,
\newblock Phys. Rev. A 66 (2002) 023609

\bibitem{timmermans98}
E.~Timmermans and R. Cote,
\newblock Phys. Rev. Lett. 80 (1998) 3419

\bibitem{miller62}
A.~Miller, D. Pines and P. Nozieres,
\newblock Phys. Rev. 127 (1962) 1452

\bibitem{chikkatur2000}
A.~P.~Chikkatur, A.~G\"orlitz, D.~M.~Stamper-Kurn, S. Inouye,
S.~Gupta and W.~ Ketterle,
\newblock Phys. Rv. Lett. 85 (2000) 483

\bibitem{roberts2006}
D.~C.~Roberts,
\newblock Phys. Rev. A 74 (2006) 013613

\bibitem{roberts2005}
D.~C.~Roberts and Y.~Pomenau,
\newblock Phys. Rev. Lett. 95 (2005) 145303

\end{thebibliography}
\end{document}